\documentclass[12pt,aps,prd,onecolumn,superscriptaddress,nofootinbib,amsmath,amssymb]{revtex4}
\usepackage{graphicx}
\usepackage{wrapfig}
 \usepackage[hypertexnames=false]{hyperref}
\begin{document}

 \title{A note on the Cardy formula and black holes in 3d (massive) bigravity}

\author{M\'{a}ximo Ba\~{n}ados}
\email{maxbanados@fis.puc.cl}
\affiliation{Departamento de F\'{\i}sica,\\
P. Universidad Cat\'{o}lica de Chile, Casilla 306, Santiago 22,Chile. }

\author{Miguel Pino}
\email{mnpino@uc.cl}
\affiliation{Departamento de F\'{\i}sica,\\
P. Universidad Cat\'{o}lica de Chile, Casilla 306, Santiago 22,Chile. }

\begin{abstract}
We report full agreement between the Bekenstein-Hawking semiclassical entropy and Cardy's formula for black holes in 3d (massive) bigravity. Since this theory does have local bulk local excitations (a massive graviton) and black holes are {\it not} locally AdS   this example provides a non-trivial test of the robustness of the Cardy formula.
\end{abstract}

\maketitle

Since the successful application of Cardy's formula \cite{Strominger,Birmingham:1998jt} to 3d black holes \cite{BTZ}, countless examples of black holes with a conformal symmetry have been discussed and exact agreement with Bekenstein-Entropy is found in all cases (see, for example, \cite{Guica:2008mu,Carlip:1999cy,Carlip:1998wz,Bousso:2001mw,
Solodukhin:1998tc,Setare:2009hw,Setare:2003fg,Setare:2003jc,Jejjala:2009if,Button:2010kg} and references therein and thereof).

The goal of this short note is to report yet another successful example, namely, three-dimensional  bigravity.  This is a non-trivial example in the sense that this theory does have propagating degrees of freedom (a massive graviton), and the black hole solutions are not locally AdS. Newton potentials, for example, have terms of the form $r^\alpha$ ($\alpha$ a negative constant). Furthermore, the conformal symmetry lives at the asymptotic $r\rightarrow\infty $ region, while the black hole properties are captured by the near horizon geometry. The validity of Cardy's formula is thus not a direct extension of \cite{Strominger,Birmingham:1998jt}.

Bigravity, as first formulated by Isham, Salam and Strathdee \cite{ISS}, is a theory for two dynamical metrics $g_{\mu\nu}$ and $f_{\mu\nu}$ with action,
\begin{eqnarray}\label{I}
I[g_{\mu\nu},f_{\mu\nu}] &=& {1 \over 16\pi G} \int \Big[\ \sqrt{g} R(g) + \sigma\, \sqrt{f} R(f)- U(g,f) \Big].
\end{eqnarray}
$U(g,f)$ is an interaction potential. The dimensionless parameter $\sigma$ measures the relative strengths of both Newton's constants. Matter can be added coupled to either metric, or both. We shall work here in vacuum. We also assume that the potential does not contain derivatives of the metric fields. A specific example will be displayed below.

The application of Cardy formula to black hole physics require three ingredients: a black hole solution with an asymptotic conformal symmetry, the relation between the zero mode Virasoro field and black hole parameters, and the corresponding Brown-Henneaux \cite{BH} central charge.

For bigravity all these ingredients can be read of from the action in a straightforward way.

First, as discussed in \cite{BT3d}, there exists solutions (for particular potentials, see below) that behave asymptotically as
\begin{eqnarray}
  dg^2 &=& {r^2 \over \ell^2} dt^2 + {\ell^2  \over r^2}dr^2 + r^2 d\phi^2 + \cdots \\
  df^2 &=& N^2 \left( {r^2 \over \ell^2} dt^2 +  {\ell^2 \over r^2}dr^2 + r^2 d\phi^2 + \cdots \right)  \label{AdSf}
\end{eqnarray}
where $N$ is a constant and $\ell$ is an AdS radius depending on the parameters in the potential ($dg^2\equiv g_{\mu\nu}dx^{\mu}dx^{\nu}$ and $df^2\equiv f_{\mu\nu}dx^{\mu}dx^{\nu}$). These solutions are asymptotically AdS, and have a conformal symmetry \cite{BT3d}\footnote{Note that both metrics are asymptotic to the same AdS space, with the same speed of light, in the same coordinate system.}. The central charge is
\begin{equation}
c = {3\ell \over 2G} + \sigma {3N\ell \over 2G}\label{c}
\end{equation}
This formula can be computed explicitly from the generators (see \cite{BT3d} for more details). It is more illuminating, however, to explain (\ref{c}) directly from the action (\ref{I}). The first term is the expected contribution from ${1 \over 16\pi G }\sqrt{g}R$. The second metric contributes in the same way. The only difference is that Newton's constant for $f_{\mu\nu}$ appears in (\ref{I}) divided by $\sigma$, and the AdS radius of (\ref{AdSf}) is not $\ell$ but $N\ell$. The total central charge is then (\ref{c}).

Second, asymptotically AdS black holes in bigravity do exist. As recently discussed in \cite{BGP,JacobsonDeffayet}, the horizon for black holes in bigravity must be located at the same spacetime point. General arguments imply that the Bekenstein-Hawking entropy for black holes in bigravity is then
\begin{equation}\label{SBH}
S = {A(g) \over 4G}  + \sigma {A(f) \over 4G}
\end{equation}
where $A(g,f)$ refers to the area of the horizon of each metric.

Finally, since the potential does not have derivatives of the fields in the action (\ref{I}), the total energy of any configuration is  simply the sum of two ADM functionals,
\begin{equation}
E = E_{ADM}(g) + \sigma E_{ADM}(f).
\end{equation}
This allows a simple calculation of the total mass, and its relation to the conformal generators.

With these ingredients at hand, we now proceed to check Cardy's formula. At this point we fix the potential so that an explicit solution is available. We consider the 2-parameter potential
\begin{eqnarray}\label{U2}
U(g,f) &=& \sqrt{f}( g_{\mu\nu}-f_{\mu\nu}) (g_{\alpha\beta}-f_{\alpha\beta} )\times \nonumber\\
&& \ \ \ \ \ \   \Big[p_1^2 ( f^{\mu\alpha} f^{\nu\beta} - f^{\mu\nu}f^{\alpha\beta}) -  p_2^2( g^{\mu\alpha}g^{\nu\beta} - g^{\mu\nu}g^{\alpha\beta})\Big].
\end{eqnarray}
where $p_1,p_2$ are arbitrary real parameters. Here $f^{\mu\nu}$ represents the inverse of $f_{\mu\nu}$.

The equations of motion following from the action (\ref{I}) are solved by the spherically symmetric, static, black holes solution with metrics \cite{Berezhiani:2008nr,BGP}
\begin{eqnarray}
dg^2&=&-hdt^2+\frac{dr^2}{h}+r^2d{\phi}^2\nonumber\\
df^2&=&-Xdt^2+Ydr^2+2Hdrdt+k^2 r^2 d{\phi}^2\label{sol}
\end{eqnarray}
where $h$, $X$, $Y$ and $H$ are given by
\begin{eqnarray}
h&=&\frac{r^2}{A^2}-8M_g-\frac{Qk^2\sigma \kappa}{\alpha}r^\alpha\\
X&=&\frac{r^2}{B^2}-8M_f+\frac{Q}{\alpha}r^\alpha\\
H&=&\sqrt{\frac{1}{\kappa^2 k^2}-XY}
\end{eqnarray}
while $Y$ is best expressed by the relation,
\begin{equation}
hY+\frac{X}{h}=\frac{Qk^2\sigma}{4(2k^2-1)p_1^2}r^{\alpha-2}+k^2+\frac{1}{\kappa^2k^4}.
\label{Y}
\end{equation}

This configuration has four integration constants: $M_g$, $M_f$, $k$ and $Q$. The constants $\kappa$, $A$ and $B$ are not independent but related to $k$ and the action parameters $p_1,p_2,\sigma$ by,
\begin{eqnarray}
\kappa&=&\frac{p_2}{p_1}\frac{\sqrt{k^2-2}}{\sqrt{1-2k^2}}\\
\frac{1}{A^2}&=&\frac{4p_1^2(k-1)(k+1)(1-k^8\kappa^2+2\kappa^2k^6-2k^2)}{\kappa k^4(k^2-2)}\\
\frac{1}{B^2}&=&\frac{p_1^2(k-1)(k+1)(4k^8\kappa^2-11\kappa^2k^6+3\kappa^2k^4+3k^4+5k^2-4)}{\kappa^2 k^6 \sigma (k^2-2)}
\end{eqnarray}
The exponent $\alpha$ is related only to $k$ by the relation,
\begin{equation}
\alpha=\frac{4k^4-7k^2+4}{(k^2-2)(2k^2-1)}
\end{equation}
The constant $k$ is arbitrary but we restrict its range by asymptotically AdS boundary conditions, i.e., demanding $\alpha$ to be negative. This implies,
\begin{equation}
 \sqrt{1/2}<k<\sqrt{2}
\label{k0range}
\end {equation}

Since $M_f$ and $M_g$ are independent, the horizons of $g_{\mu\nu}$ and $f_{\mu\nu}$ are located at different spacetime points. However, as discussed in \cite{BGP,JacobsonDeffayet}, each metric is singular at the horizon of the other. Thus, the only way to have a smooth and regular solution is to choose $M_g$ and $M_f$ such that both horizons are located at the same point. Let $r=r_0$ the location of the horizon. $M_g$ and $M_f$ are then given by
\begin{eqnarray}
8M_g&=& \frac{r_0^2}{A^2}-\frac{Qk^2\sigma \kappa}{\alpha}r_0^{\alpha} \nonumber\\
8M_f&=&  \frac{r_0^2}{B^2}+\frac{Q}{\alpha}r_0^{\alpha} \label{masas}
\end{eqnarray}

We now explore the asymptotic behavior at $r\rightarrow \infty$. Assuming (\ref{k0range}) is valid, the leading terms in both metrics are,
\begin{eqnarray}
dg^2&\approx&-\frac{r^2}{A^2}dt^2+\frac{A^2}{r^2}dr^2+r^2d\phi^2\nonumber\\
df^2&\approx&-\frac{r^2}{B^2}dt^2+\frac{C}{r^2}dr^2+\frac{2}{k}\sqrt{\left(k^2-\frac{A^2}{B^2}\right)
\left(\frac{1}{\kappa^2k^2}-\frac{A^2k^2}{B^2}\right)}drdt+k^2r^2d\phi^2\label{asym}
\end{eqnarray}
where
\begin{equation}
C=A^2\left( k^2+\frac{1}{\kappa^2k^4}-\frac{A^2}{B^2}\right)
\end{equation}
We demand both metrics to be proportional to some AdS background
\begin{equation}
ds^2=-\frac{r^2}{\ell^2}dt^2+\frac{\ell^2}{r^2}dr^2+r^2d\phi^2\label{ads}
\end{equation}
for some constant AdS radius $\ell$. Directly from (\ref{ads}) and (\ref{asym}) one concludes that the  only way to accomplish this condition is  by imposing two conditions,
\begin{eqnarray}
k^2&=&\frac{A^2}{B^2}\label{rel1}\\
\frac{B^2C}{A^2}&=&A^2\label{rel2}
\end{eqnarray}
These conditions\footnote{It is useful to note that relation (\ref{rel2}) can be written as: $\frac{1}{\kappa}=k^3\label{rel3}$} put constraints on the action parameters and the value of $k$. But they do not restrict the horizon location which is still arbitrary, as it must be.

Summarizing, if relations (\ref{rel1}) and (\ref{rel2}) hold, then both metrics are asymptotically AdS in the same coordinate system and the constant $N$ entering in the central charge (\ref{c}) is equal to,
\begin{equation}\label{N}
N = {A \over B}.
\end{equation}

Just as in the purely gravitational case \cite{BH}, fluctuations around the AdS backgrounds are characterized by Virasoro fields with a non-zero classical central charge (\ref{c}) (see \cite{BT3d} for details). The last step to apply the Cardy formula is to express the zero mode Virasoro charges $L_0,\bar L_0$ is terms of the black hole parameters. This is a straightforward, albeit long, calculation that we omit. The result is,
\begin{equation}\label{L0}
L_0=\bar{L}_0=\frac{1}{2G}\big( A M_g +\sigma B M_f \big)
\end{equation}
Note that these solutions are non-rotating hence $L_0=\bar L_0$.

We are now ready to apply Cardy formula,
\begin{equation}
S=2\pi\sqrt{\frac{cL_0}{6}}+2\pi\sqrt{\frac{c\bar{L}_0}{6}}\label{cardy}
\end{equation}
for the density of states. Plugging the central charge (\ref{c}), the zero modes (\ref{L0}) and using  (\ref{masas}), (\ref{rel1}) and (\ref{rel2}), we obtain,
\begin{equation}
S=\frac{2\pi r_0}{4G}(1+\sigma k),
\end{equation}
exactly equal to one fourth of the sum of both horizon areas (see (\ref{SBH})), as promised.

\acknowledgements

We would like to thank A. Gomberoff for many illuminating discussions on the Cardy formula. MB was partially supported by Fondecyt (Chile) Grants \#1100282 and \# 1090753. MP was supported by CONICYT grant (Chile) and VRAID grant (PUC, Chile).

\bibliographystyle{abbrv}
\bibliography{/home/mnpino/Desktop/Papers/Bib.bib}

\begin{thebibliography}{10}

\bibitem{BTZ}
M.~Ba\~nados, C.~Teitelboim, and J.~Zanelli.
\newblock {The Black hole in three-dimensional space-time}.
\newblock {\em Phys. Rev. Lett.}, 69:1849--1851, 1992.

\bibitem{BT3d}
M.~Ba\~nados and S.~Theisen.
\newblock {Three-dimensional massive gravity and the bigravity black hole}.
\newblock {\em JHEP}, 11:033, 2009.

\bibitem{BGP}
M.~Banados, A.~Gomberoff, and M.~Pino.
\newblock {The bigravity black hole and its thermodynamics}.
\newblock 2011.
\newblock * Temporary entry *.

\bibitem{Berezhiani:2008nr}
Z.~Berezhiani, D.~Comelli, F.~Nesti, and L.~Pilo.
\newblock {Exact Spherically Symmetric Solutions in Massive Gravity}.
\newblock {\em JHEP}, 07:130, 2008.

\bibitem{Birmingham:1998jt}
D.~Birmingham, I.~Sachs, and S.~Sen.
\newblock {Entropy of three-dimensional black holes in string theory}.
\newblock {\em Phys. Lett.}, B424:275--280, 1998.

\bibitem{Bousso:2001mw}
R.~Bousso, A.~Maloney, and A.~Strominger.
\newblock {Conformal vacua and entropy in de Sitter space}.
\newblock {\em Phys.Rev.}, D65:104039, 2002.

\bibitem{BH}
J.~D. Brown and M.~Henneaux.
\newblock {Central Charges in the Canonical Realization of Asymptotic
  Symmetries: An Example from Three-Dimensional Gravity}.
\newblock {\em Commun. Math. Phys.}, 104:207--226, 1986.

\bibitem{Button:2010kg}
B.~K. Button, L.~Rodriguez, and C.~A. Whiting.
\newblock {A Near Horizon CFT Dual for Kerr-Newman-$AdS$}.
\newblock {\em Int.J.Mod.Phys.}, A26:3077--3090, 2011.

\bibitem{Carlip:1998wz}
S.~Carlip.
\newblock {Black hole entropy from conformal field theory in any dimension}.
\newblock {\em Phys.Rev.Lett.}, 82:2828--2831, 1999.

\bibitem{Carlip:1999cy}
S.~Carlip.
\newblock {Entropy from conformal field theory at Killing horizons}.
\newblock {\em Class.Quant.Grav.}, 16:3327--3348, 1999.

\bibitem{JacobsonDeffayet}
C.~Deffayet and T.~Jacobson.
\newblock {On horizon structure of bimetric spacetimes}.
\newblock 2011.
\newblock * Temporary entry *.

\bibitem{Guica:2008mu}
M.~Guica, T.~Hartman, W.~Song, and A.~Strominger.
\newblock {The Kerr/CFT Correspondence}.
\newblock 2008.

\bibitem{ISS}
C.~J. Isham, A.~Salam, and J.~A. Strathdee.
\newblock {F-dominance of gravity}.
\newblock {\em Phys. Rev.}, D3:867--873, 1971.

\bibitem{Jejjala:2009if}
V.~Jejjala and S.~Nampuri.
\newblock {Cardy and Kerr}.
\newblock {\em JHEP}, 1002:088, 2010.

\bibitem{Setare:2003jc}
M.~Setare and M.~Altaie.
\newblock {The Cardy-Verlinde formula and entropy of topological Kerr-Newman
  black holes in de Sitter spaces}.
\newblock 2003.

\bibitem{Setare:2009hw}
M.~Setare and M.~Jamil.
\newblock {The Cardy-Verlinde Formula and Entropy of the Charged Rotating BTZ
  black Hole}.
\newblock {\em Phys.Lett.}, B681:469--471, 2009.

\bibitem{Setare:2003fg}
M.~R. Setare and E.~C. Vagenas.
\newblock {Cardy-Verlinde formula and Achucarro-Ortiz black hole}.
\newblock {\em Phys.Rev.}, D68:064014, 2003.

\bibitem{Solodukhin:1998tc}
S.~N. Solodukhin.
\newblock {Conformal description of horizon's states}.
\newblock {\em Phys.Lett.}, B454:213--222, 1999.

\bibitem{Strominger}
A.~Strominger.
\newblock {Black hole entropy from near-horizon microstates}.
\newblock {\em JHEP}, 02:009, 1998.

\end{thebibliography}
\end{document}